\documentclass[conference]{IEEEtran}
\IEEEoverridecommandlockouts
\usepackage{multirow}
\usepackage{cite}
\usepackage{amsmath,amssymb,amsfonts}
\usepackage{algorithmic}
\usepackage{graphicx}
\usepackage{textcomp}
\usepackage{xcolor}
\usepackage{tikz}
\usepackage{hyperref}
\usepackage{fancyhdr}

\def\BibTeX{{\rm B\kern-.05em{\sc i\kern-.025em b}\kern-.08em
    T\kern-.1667em\lower.7ex\hbox{E}\kern-.125emX}}

\makeatletter
\def\ps@IEEEtitlepagestyle{%
  \def\@oddfoot{\mycopyrightnotice}%
  \def\@evenfoot{}%
}
\def\mycopyrightnotice{%
  {\footnotesize \begin{minipage}{\textwidth}\ \\[12pt] \centering \textcopyright 2021 IEEE.  Personal use of this material is permitted.  Permission from IEEE must be obtained for all other uses, in any current or future media, including reprinting/republishing this material for advertising or promotional purposes, creating new collective works, for resale or redistribution to servers or lists, or reuse of any copyrighted component of this work in other works.\hfill \end{minipage}}
  \gdef\mycopyrightnotice{}
}

\makeatother

\begin{document}
\title{Password authentication schemes \\on a quantum computer}
\thispagestyle{plain}
\pagestyle{plain}
\author{\IEEEauthorblockN{Sherry Wang}
\IEEEauthorblockA{\textit{School of Electrical Engineering}\\ \textit{and Computer Science,} \\
\textit{University of Ottawa,}\\
Ottawa, ON, Canada \\
swang261@uottawa.ca}
\and
\IEEEauthorblockN{Carlisle Adams}
\IEEEauthorblockA{\textit{School of Electrical Engineering}\\ \textit{and Computer Science,} \\
\textit{University of Ottawa,}\\
Ottawa, ON, Canada \\
cadams@uottawa.ca}
\and
\IEEEauthorblockN{Anne Broadbent}
\IEEEauthorblockA{\textit{Department of Mathematics}\\
\textit{and Statistics,} \\
\textit{University of Ottawa,}\\
Ottawa, ON, Canada \\
abroadbe@uottawa.ca}
}

\maketitle

\begin{abstract}
In a post-quantum world, where attackers may have access to full-scale quantum computers, all classical password-based authentication schemes will be compromised. Quantum copy-protection prevents adversaries from making copies of existing quantum software; we suggest this as a possible approach for designing post-quantum-secure password authentication systems. In this paper, we show an implementation of quantum copy-protection for password verification on IBM quantum computers. We also share our quantum computation results and analyses, as well as lessons learned.
\end{abstract}


\section{Introduction}
Full-scale quantum computers will be able to achieve quadratic speed-ups in search problems compared with the classical computers that are in use today. Grover’s algorithm is one such quantum algorithm. Given a particular output value of a black box function, Grover’s algorithm finds the unique input in O($\sqrt{n}$) evaluations \cite{b1}. Classical algorithms on the other hand use O(\textit{n}) evaluations. Thus, Grover’s algorithm could brute-force a 256-bit key in approximately 2\textsuperscript{128} operations.

The implication of Grover’s algorithm on passwords is that all classical password-based authentication will be vulnerable to attack. Grover’s algorithm could be used to invert a cryptographic hash function, allowing attackers to find a victim’s password far more quickly than is possible today. In addition, users sometimes choose weak passwords, which has facilitated countless security breaches over the years. Users would have to create and remember good quality passwords that are at least four times longer than what is used today in a post-quantum environment. However, it is unreasonable to expect this of the average user. An alternate strategy is to design systems in which offline password guessing attacks are harder or even impossible for attackers to mount. In our work, we explore the use of quantum copy-protection of point functions for password verification in a post-quantum environment. We implement password authentication schemes on IBM quantum computers using Python, Qiskit, and liboqs-python\footnote{Our code is available at \href{https://github.com/wsherry/QuantumPasswordAuthentication}{https://github.com/wsherry/QuantumPassword\\Authentication}}.

The ‘no cloning’ theorem gives quantum information one of its most interesting properties---it cannot be copied. Quantum copy-protection makes use of the theorem to construct programs that cannot be copied or pirated. In our work, we are interested in quantum copy-protection of point functions based on the recent work of Broadbent, Jeffery, Lord, Podder, and Sundaram \cite{b2}. Point functions are special functions that map \textit{n}-bit strings to a 1-bit string. 
The output is 1 only if the input matches the point \textit{p} of the point function, otherwise it is~$0$. The point \textit{p} in the point function is the key used to encode quantum copy-protected programs. Quantum copy-protection of point functions translates well into a password verification scheme, since its point \textit{p} can be thought of as the password, and only the correct password will provide the outcome we want. 

We test our implementation using the Honest-Malicious security model introduced in \cite{b2}. For a correctness check of our password authentication scheme, we use a simplified version of the game in \cite{b2}. The game occurs between a challenger and an evaluator. The challenger prepares a copy-protected program using \textit{p} as the key and sends it to an evaluator. The challenger then prepares a challenge input sampled from the challenge distribution given in \cite{b2}. This challenge distribution outputs:
\begin{itemize}
  \item with probability ½, \textit{p}; and
  \item with probability ½, a uniformly sampled \textit{n}-bit string \\ \textit{x} $\leftarrow$ \{0,1\}\textsuperscript{n} \textbackslash \{\textit{p}\}. 
\end{itemize}
The correctness is checked at the end when the evaluator evaluates the challenge inputs on the program, and the error syndromes are checked.

\begin{figure}[htbp]
\resizebox{\columnwidth}{!}
{
\includegraphics{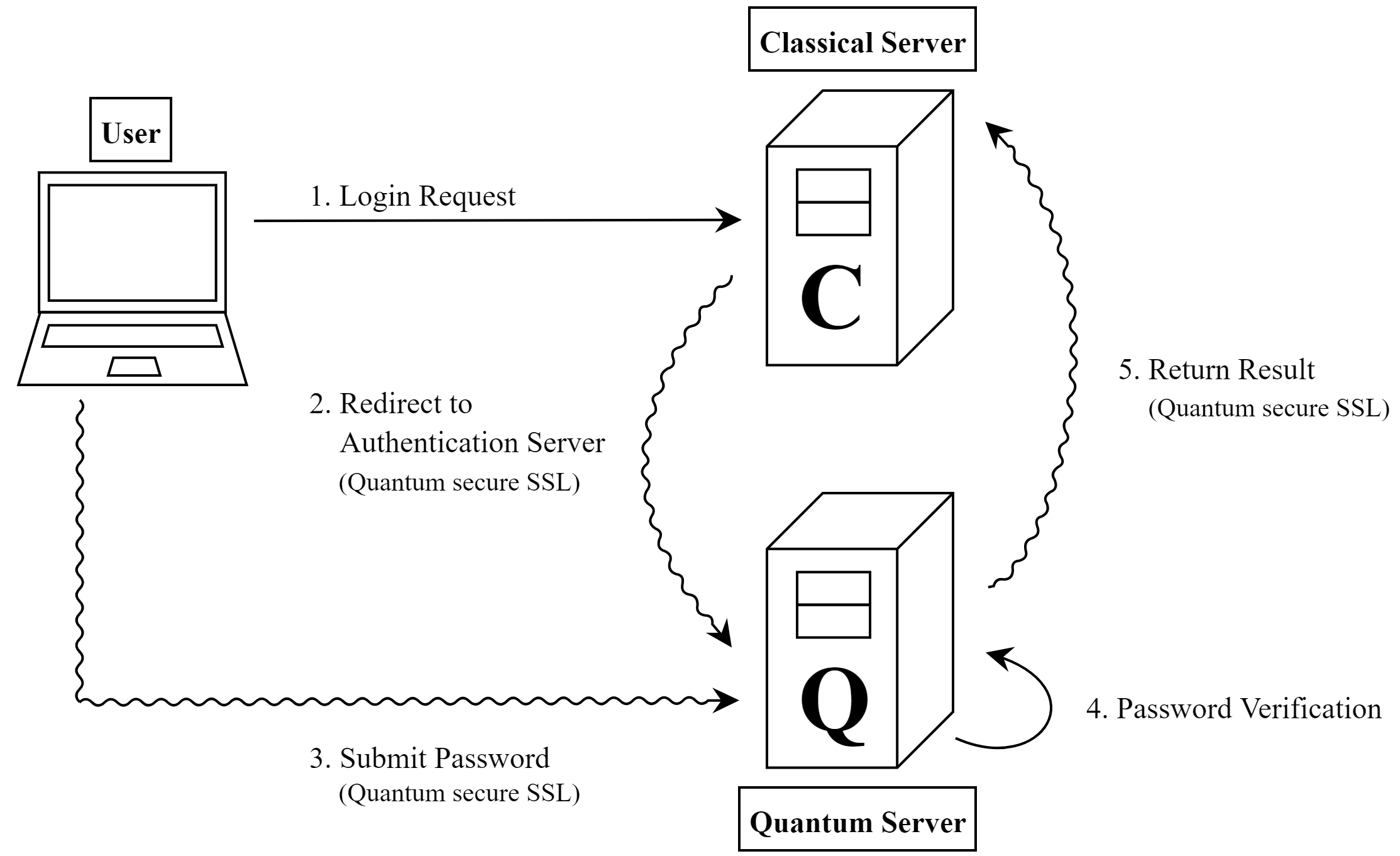}
}
\caption{The proposed password authentication model. The wavy arrows indicate the use of quantum secure SSL.}
\label{fig}
\end{figure}
In addition to implementing password authentication schemes, we present a model of how these password authentication schemes interact with users and applications in a post-quantum world (see Fig. 1). The model consists of various different systems and actors—--a classical server on which the main application is hosted, a quantum authentication server for password verification, and users. We made the assumption that applications would still be held on classical servers. This assumption allowed us to compartmentalize quantum functions, minimizing the complexity and depth of the circuits run on quantum computers. Users who want to access applications that require a login will first access the classical servers on which these applications will be hosted. When users want to log in to their respective accounts, the classical server will reroute the user to a quantum authentication server. Password verification occurs, and the user is redirected according to the results of the password verification. In the game defined earlier, the challenger represents the classical and authentication servers, the evaluator represents the user, and the challenge inputs can be thought of as the attempted passwords.

To ensure that the communication is secure between the different actors and systems, we  implement a quantum secure version of Secure Sockets Layer (SSL). We use FrodoKEM-1344 for the initial key exchange \cite{b3}, followed by the Advanced Encryption Standard 256 (AES-256) algorithm for secure password exchange \cite{b4}. FrodoKEM-1344 and AES-256 are schemes which are considered to be quantum-resistant.

There has been some related research on implementing security methodologies on the quantum cloud, including the implementations of homomorphic encryption schemes \cite{b5}, quantum locker protocols for message retrieval \cite{b6}, and secret sharing protocols \cite{b7}. However, there seems to be little publication on password authentication schemes on quantum computers, and our work may be seen as an early step to fill the gap.
\section{Implementation}

The authentication server is a quantum computer that uses copy-protection of point functions to verify passwords. To construct a copy-protected program, we implement a total quantum authentication scheme that uses strong trap codes, based on the work in \cite{b8}. The quantum authentication scheme starts with two traps in the states $\vert$0$\rangle$ and $\vert$+$\rangle$, respectively, to which six auxiliary qubits (in the state $\vert$0$\rangle$) are appended to each trap. Each trap qubit and its corresponding six auxiliary qubits are then encoded using a [7,1,3] quantum error correction code---the Steane code, based on the implementation in \cite{b9}. All the qubits are then permuted and encrypted using the quantum one-time pad. In our implementation, to reduce noise, the quantum one-time pad and the permutation are performed classically. Using the principles of delegated computing in \cite{b10}, we are able to move the one-time pad to the beginning of the circuit. Meanwhile, the permutation determines how the CNOTs in the Steane code are mapped. These design choices were guided by experimentation after running the program for several iterations.

The encoded password program seen in Fig. 2 serves as the copy-protected program. For the verification step, a challenge input can be used to decrypt the password program (see Fig.~3). Our circuit only detects phase flip errors, and the error syndrome is extracted from the circuit using six auxiliary qubits.

To further mitigate the effects of quantum computer noise, we directly take the error syndrome from the password program, and we do all the necessary processing and decrypting classically.

\begin{figure}[htbp]
\resizebox{\columnwidth}{!}{
\includegraphics{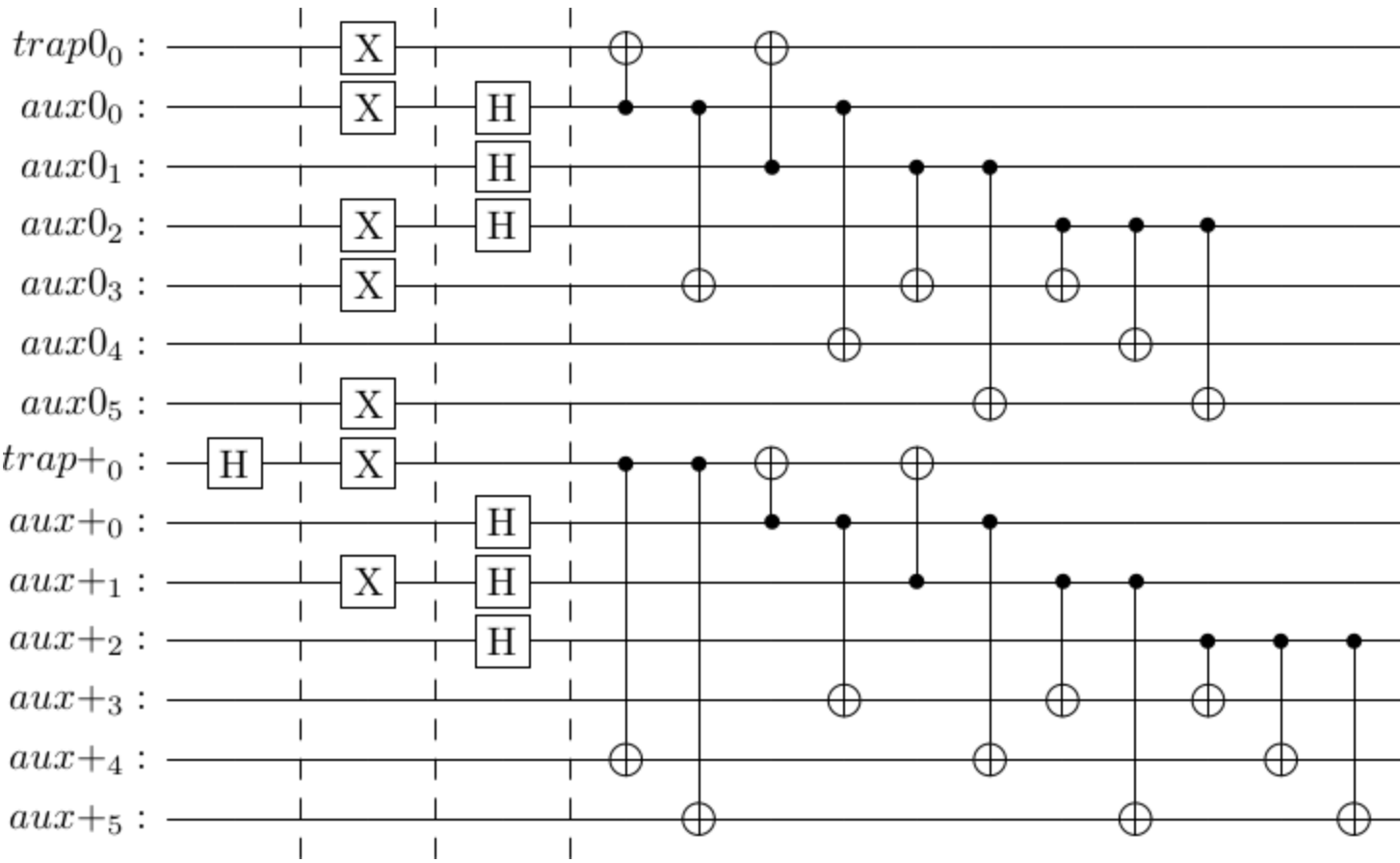}
}
\caption{ Password program prepared in the authentication server. Starting from the left, the first section is the initialization of the trap qubits. The second section is the one-time pad encryption, and the third and fourth sections are two Steane codes that have an identity permutation.}
\label{fig}
\end{figure}

\begin{figure*}
\includegraphics[width=\textwidth,height=\textheight,keepaspectratio]{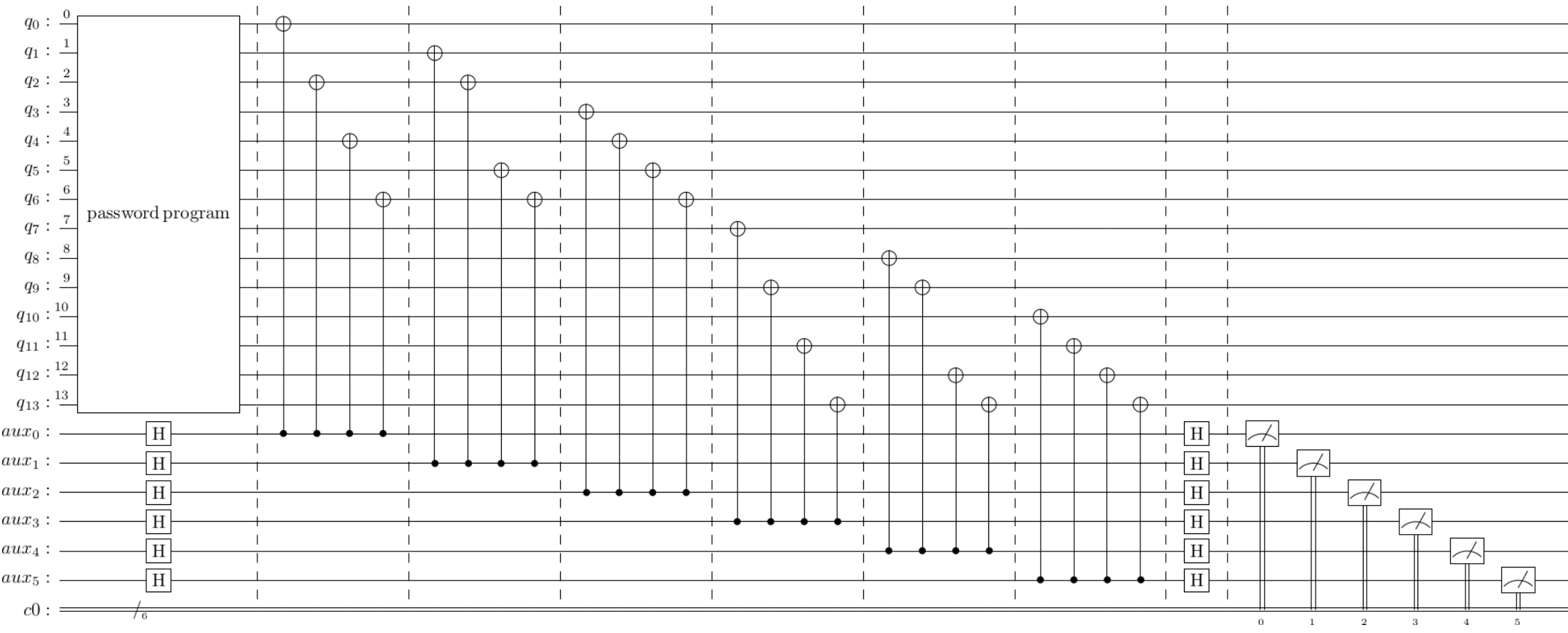}
\caption{The verification circuit used to extract the error syndrome measurement of the password program (shown in Fig. 2) for phase flip errors only.}
\end{figure*}

\begin{figure*}
\resizebox{\columnwidth}{!}
{
\includegraphics{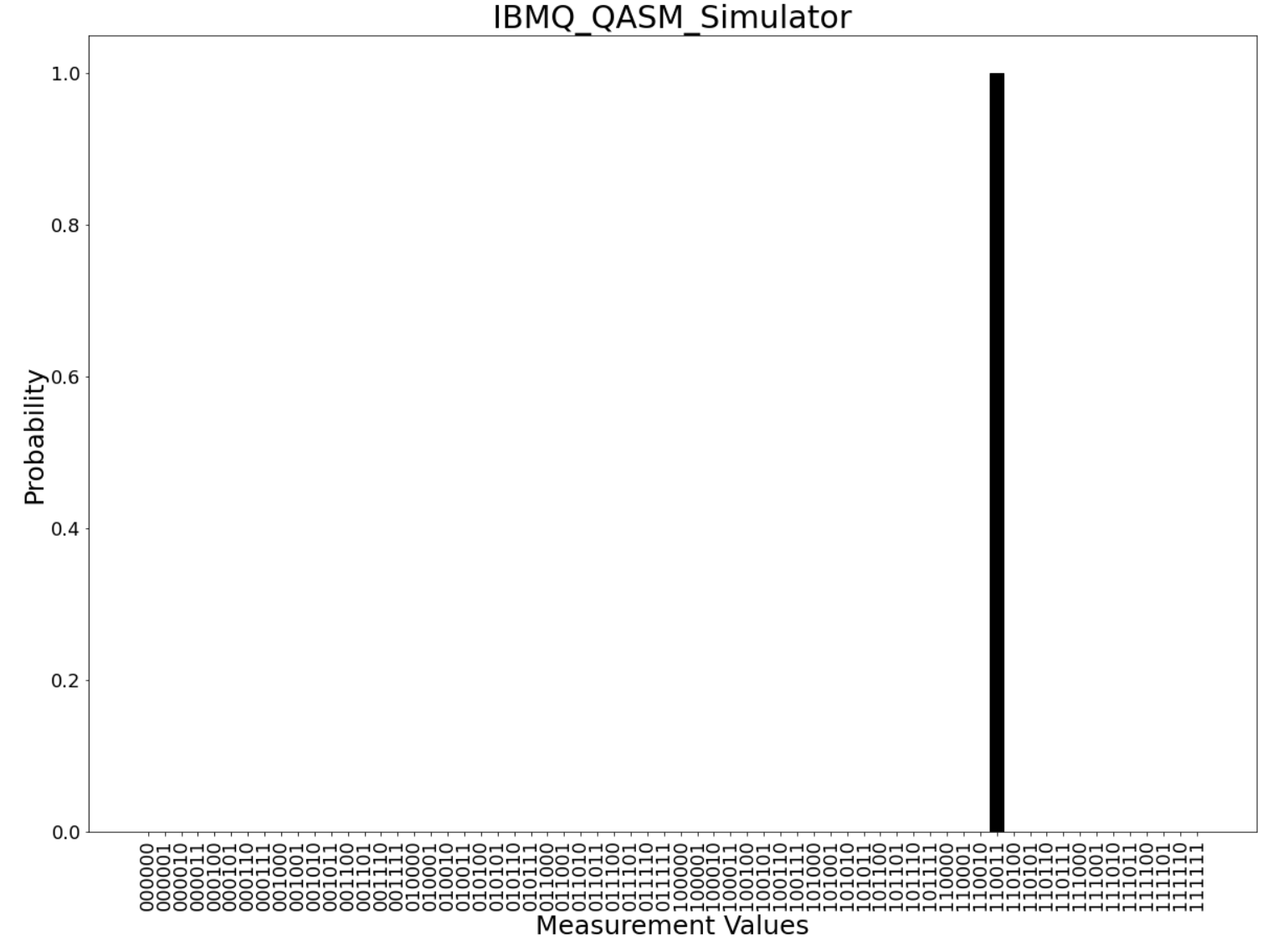}
}
\resizebox{\columnwidth}{!}
{
\includegraphics{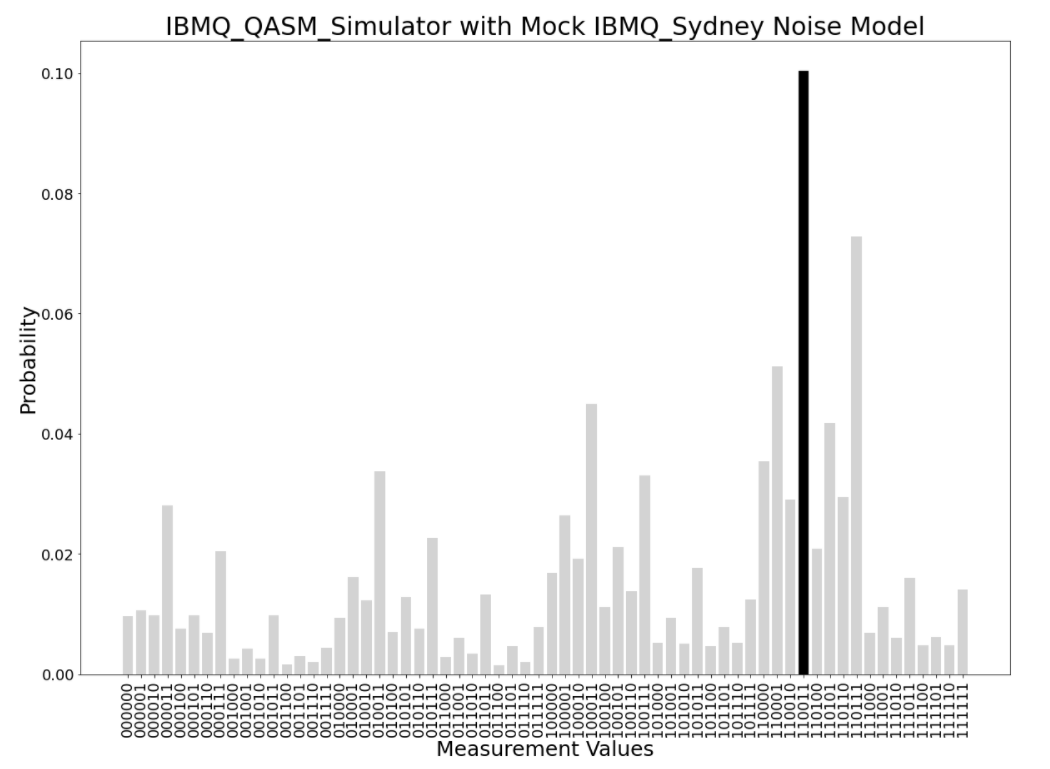}
}
\par 
\resizebox{\columnwidth}{!}
{
\includegraphics{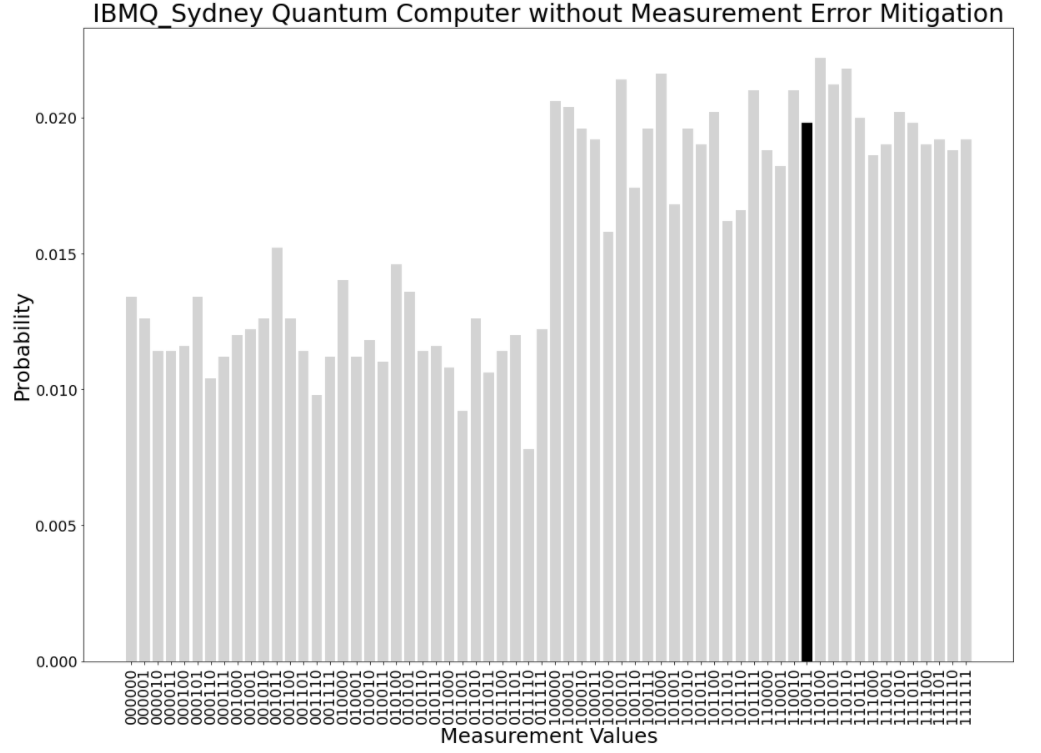}
}
\resizebox{\columnwidth}{!}
{
\includegraphics{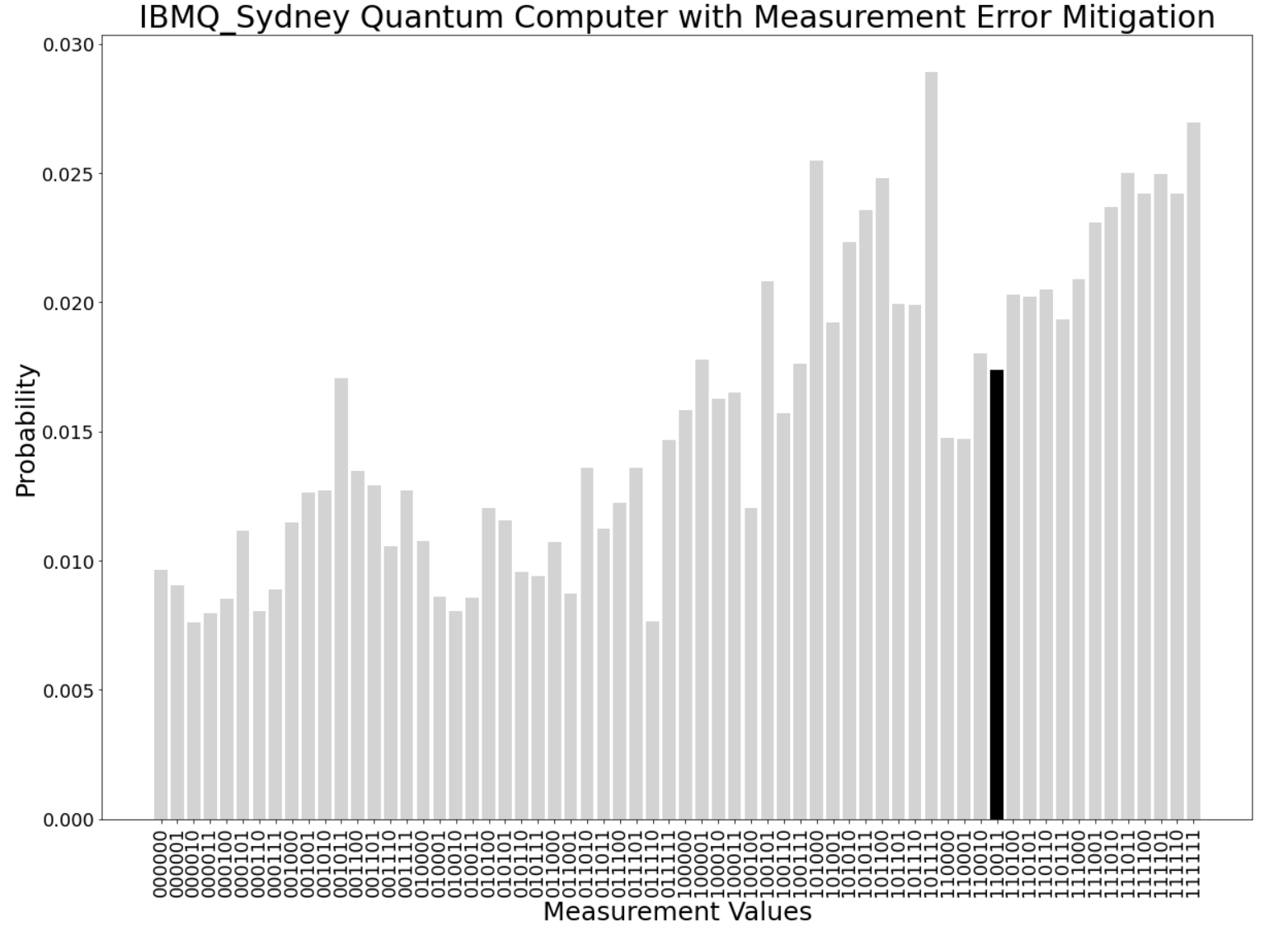}
}
\par
\caption{The error syndrome measurements of the same copy-protected program for test one. The program is run on the ibmq\_qasm\_simulator (top left), ibmq\_qasm\_simulator with a mock ibmq\_sydney device noise model (top right), the ibmq\_sydney quantum device without Qiskit’s measurement error mitigation (bottom left), and the ibmq\_sydney device with all error mitigation techniques (bottom right).The dark bar in each graph represents the expected measurement outcome, which when post-processed equates to the no error syndrome (all 0s). Each circuit was executed and sampled 5000 times.}
\end{figure*}
For the quantum secure SSL, we used Open Quantum Safe’s liboqs-python for implementing FrodoKEM-1344, and PyCryptodome for implementing AES-256. 
\section{Results}
We ran five different tests for correctness on IBM's 27-qubit quantum computer, ibmq\_sydney. For the first test, we used a challenge input that was the same as the point \textit{p} used in the program. The second to fifth tests used challenge inputs that were different from \textit{p} to simulate an attack on the program. The second test has a permutation error, while the third and fourth tests have X and Z errors, respectively. The fifth test checks for correctness when both X and Z errors are introduced to the circuit. Finally, a sixth test takes a challenge point from the challenge distribution defined earlier. 

The error syndromes are measured and post-processed classically to determine true positive rates for tests one and six, and the false positive rates for tests two to six.
The graphs for the first test where the challenge input is equal to \textit{p}, are shown in Fig. 4. Typically, the error syndromes measured should produce all 0s if there are no errors, however since we delay the one-time pad decryption and inverse permutation and perform it classically after measuring, the results seen in Fig.~4 are the raw measurements. The same \textit{p} was run across two simulators, and an actual quantum device. The ideal simulator provides the ideal scenario: the expected result occurs 100\% of the time. When we place an ibmq\_sydney noise model on the simulator, noise can be seen, and the expected result only occurs about 10\% of the time.

Finally, when the program is run on the actual device, we see that there is much more noise than what the noise model had covered. The averaged true and false positive rates for each test can be seen in Table 1.

\begin{table}[hbtp]
\caption{The average true and false positive rates for the correctness tests from the ibmq\_sydney quantum computer, where only phase errors are detected. Each test size is 10.}
\begin{center}
\begin{tabular}{lllll}
\hline
\multicolumn{3}{|c|}{\textbf{Test}} & \multicolumn{2}{c|}{\textbf{\begin{tabular}[c]{@{}c@{}}Average Program Results \\ (ibmq\_sydney)\end{tabular}}} \\ \hline
\multicolumn{1}{|c|}{\textit{\textbf{\#}}} & \multicolumn{2}{c|}{\textit{\textbf{Description}}} & \multicolumn{1}{c|}{\textit{\textbf{\begin{tabular}[c]{@{}c@{}}True Positive \\ (Correct \\ password \\ is accepted)\end{tabular}}}} & \multicolumn{1}{c|}{\textit{\textbf{\begin{tabular}[c]{@{}c@{}}False Positive\\ (Wrong \\ password \\ is accepted)\end{tabular}}}} \\ \hline
\multicolumn{1}{|l|}{1} & \multicolumn{2}{l|}{\begin{tabular}[c]{@{}l@{}}Challenge input \\ is equal to point\end{tabular}} & \multicolumn{1}{l|}{1.42\%}  & \multicolumn{1}{l|}{-} \\ \hline
\multicolumn{1}{|l|}{2} & \multicolumn{2}{l|}{\begin{tabular}[c]{@{}l@{}}Challenge input \\ has a permutation error\end{tabular}} & \multicolumn{1}{l|}{-} & \multicolumn{1}{l|}{1.47\%} \\ \hline
\multicolumn{1}{|l|}{3} & \multicolumn{2}{l|}{\begin{tabular}[c]{@{}l@{}}Challenge input \\ has an X error\end{tabular}} & \multicolumn{1}{l|}{-} & \multicolumn{1}{l|}{1.54\%} \\ \hline
\multicolumn{1}{|l|}{4} & \multicolumn{2}{l|}{\begin{tabular}[c]{@{}l@{}}Challenge input \\ has a Z error\end{tabular}} & \multicolumn{1}{l|}{-} & \multicolumn{1}{l|}{1.88\%} \\ \hline
\multicolumn{1}{|l|}{5} & \multicolumn{2}{l|}{\begin{tabular}[c]{@{}l@{}}Challenge input \\ has an X and Z error\end{tabular}} & \multicolumn{1}{l|}{-} & \multicolumn{1}{l|}{1.79\%} \\ \hline
\multicolumn{1}{|c|}{\multirow{2}{*}{6}} & \multicolumn{1}{l|}{\multirow{2}{*}{\begin{tabular}[c]{@{}l@{}}Challenge \\ input \\ sampled\\ from \\ challenge \\ distribution\end{tabular}}} & \multicolumn{1}{l|}{\begin{tabular}[c]{@{}l@{}}Challenge \\ input \\ sampled \\ equals to  point \\ (1/10 tests)\end{tabular}} & \multicolumn{1}{l|}{1.07\%} & \multicolumn{1}{l|}{-} \\ \cline{3-5} 
\multicolumn{1}{|c|}{} & \multicolumn{1}{l|}{} & \multicolumn{1}{l|}{\begin{tabular}[c]{@{}l@{}}Challenge \\ input \\ sampled \\ does not equal \\ to point\\  (9/10 tests)\end{tabular}} & \multicolumn{1}{l|}{-} & \multicolumn{1}{l|}{1.25\%} \\ \hline
\end{tabular}
\end{center}
\end{table}

The results of the verification circuits that detect bit and phase flip errors are included in the~Appendix.

\section{Data Analysis}
The data from the quantum computer results are still very noisy. While the noise model of the ibmq\_sydney quantum computer was used, it does not completely reflect the amount of noise from the actual quantum computer, as can be seen in the graphs. There are spikes in the bar graph that are visible in the top right graph in Fig.~4. These spikes in the graph can be attributable to the fact that the error syndrome picks up phase flip errors for one of the trap codes, but due to noise, fails to correctly output the phase flip error syndrome for the other trap code. The actual ibmq\_sydney quantum computer’s results meanwhile look closer to uniform. While there are still spikes in the actual quantum device’s bar graph, they are much more subdued and rounded out. Using Qiskit’s measurement mitigation functions helps reduce some of the noise (see bottom graphs of Fig. 4). However, the noise mitigation actually increases the frequency of unexpected measurement outcomes.

\section{Lessons Learned}
Quantum computers are still far from being fully fault tolerant, and thus, are subject to much noise. As a consequence, applying research in practice may sometimes transfer poorly and produce outcomes that are noise heavy, especially if the implemented circuit has many gate operations. Using a combination of Qiskit’s error mitigation tools and performing parts of the circuit classically without compromising security can offset some of the noise.

Unlike their quantum device simulators, IBM quantum computers have a specific device layout that dictates how its virtual qubits are connected. Thus, mapping the physical qubits of the program to the actual device’s virtual qubits is an important step in mitigating noise. Qubits that have many CNOT operations with other qubits were often mapped to virtual qubits on the quantum machine that were well connected. In addition to looking at how connected each qubit is, there are CNOT operation errors between virtual qubits and single qubit Pauli-X errors that need to be taken into consideration as well. The mapping of the qubits and circuit optimizations can be done through Qiskit’s transpile function. Qiskit also offers some built-in tools for measurement noise mitigation, which can be used after executing the circuits. Using Qiskit's mock backends can provide some insight to the noise model of quantum computers which can inform decisions on where to optimize the circuit. Since quantum computers are calibrated once every 24 hour period, many of its system properties are updated to these calibrations\cite{b11}. Thus, using a noise model during simulation runs that dynamically updates according to the system properties of each quantum computer calibration, will be even more effective in reflecting the noise of the quantum machine.

High depth circuits are much more susceptible to noise. As we made modifications and gate reductions in our circuit, we were able to get slightly better results. By processing parts of the circuits classically before and after the quantum computation, we were able to reduce the depth of our circuits, and get better results. Selecting which parts of a circuit can be computed classically (without compromising security) is a very effective, but sometimes non-trivial, step in reducing noise. With the original circuit implemented, we had a circuit depth of over 200 operations; however, with some reductions including performing the one-time pad at the beginning using delegation techniques in \cite{b10} and permuting the Steane code classically, we were able to halve the depth to the 80 to 100s range. These classical operations removed many CNOTs and swap operations. 

There are other existing error mitigation techniques that can be used such as zero-noise extrapolation, which does not require additional quantum hardware resources. Mitiq is a Python toolkit that offers zero-noise extrapolation among other techniques. For more advanced error mitigation techniques, Qiskit Pulse offers qubit calibration toolkits.

\section{Conclusion}
Our work on implementing password authentication schemes can be considered as a proof of concept. The results are noisy but there are various lessons and techniques we learned during this experience to help with noise mitigation. We feel that this is a promising technique for secure password-based authentication in a post-quantum world.

\section*{Appendix}

The verification circuit for detecting both bit and phase flip errors can be seen in Fig. 5. The top 14 qubits are used for detecting bit flips, and the bottom six qubits are used for detecting phase flips.

\begin{figure*}
\includegraphics[width=\textwidth,height=\textheight,keepaspectratio]{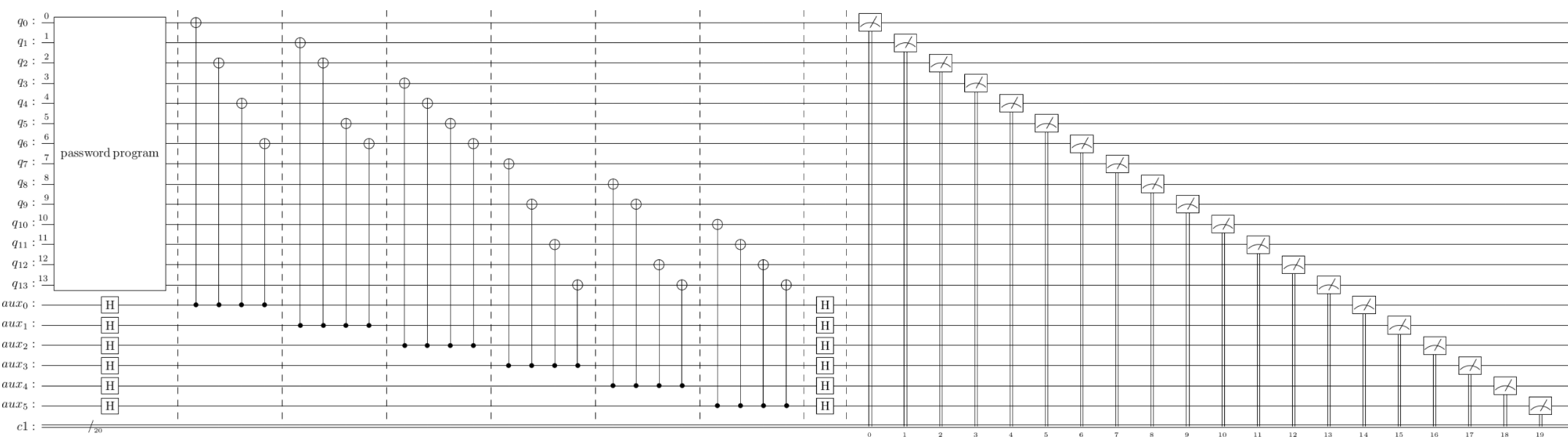}
\caption{The verification circuit used to check for bit flip errors and extract the error syndrome measurement of a password program for phase flip errors.}
\end{figure*}

We ran the five correctness tests for the bit and phase error detection verification circuits. The tests were run on IBM's 27-qubit, quantum volume 128 machine, ibmq\_montreal. The averaged true and false positive rates for each test can be seen in Tables II to IV. The true and false positive rates for detecting only bit or phase flip errors are similar in value to the ibmq\_sydney results. However, the true and false positive rates for detecting all errors, whether bit or phase flip errors, are significantly lower.

The error syndrome measurements of the same copy protected program can be seen in Fig.~6 to Fig.~8.

\begin{table}[htbp]
\caption{The average true and false positive rates for the correctness tests from the ibmq\_montreal quantum computer, where only bit flip errors are detected. Each test size is 10.}
\begin{center}
\begin{tabular}{lllll}
\hline
\multicolumn{3}{|c|}{\textbf{Test}} & \multicolumn{2}{c|}{\textbf{\begin{tabular}[c]{@{}c@{}}Average Program Results for\\ Bit Flip Detection \\ (ibmq\_montreal) \end{tabular}}} \\ \hline
\multicolumn{1}{|c|}{\textit{\textbf{\#}}} & \multicolumn{2}{c|}{\textit{\textbf{Description}}} & \multicolumn{1}{c|}{\textit{\textbf{\begin{tabular}[c]{@{}c@{}}True Positive \\ (Correct \\ password \\ is accepted)\end{tabular}}}} & \multicolumn{1}{c|}{\textit{\textbf{\begin{tabular}[c]{@{}c@{}}False Positive\\ (Wrong \\ password \\ is accepted)\end{tabular}}}} \\ \hline
\multicolumn{1}{|l|}{1} & \multicolumn{2}{l|}{\begin{tabular}[c]{@{}l@{}}Challenge input \\ is equal to point\end{tabular}} & \multicolumn{1}{l|}{2.67\%}  & \multicolumn{1}{l|}{-} \\ \hline
\multicolumn{1}{|l|}{2} & \multicolumn{2}{l|}{\begin{tabular}[c]{@{}l@{}}Challenge input \\ has a permutation error\end{tabular}} & \multicolumn{1}{l|}{-} & \multicolumn{1}{l|}{1.61\%} \\ \hline
\multicolumn{1}{|l|}{3} & \multicolumn{2}{l|}{\begin{tabular}[c]{@{}l@{}}Challenge input \\ has an X error\end{tabular}} & \multicolumn{1}{l|}{-} & \multicolumn{1}{l|}{1.61\%} \\ \hline
\multicolumn{1}{|l|}{4} & \multicolumn{2}{l|}{\begin{tabular}[c]{@{}l@{}}Challenge input \\ has a Z error\end{tabular}} & \multicolumn{1}{l|}{-} & \multicolumn{1}{l|}{2.60\%} \\ \hline
\multicolumn{1}{|l|}{5} & \multicolumn{2}{l|}{\begin{tabular}[c]{@{}l@{}}Challenge input \\ has an X and Z error\end{tabular}} & \multicolumn{1}{l|}{-} & \multicolumn{1}{l|}{1.61\%} \\ \hline
\multicolumn{1}{|c|}{\multirow{2}{*}{6}} & \multicolumn{1}{l|}{\multirow{2}{*}{\begin{tabular}[c]{@{}l@{}}Challenge \\ input \\ sampled\\ from \\ challenge \\ distribution\end{tabular}}} & \multicolumn{1}{l|}{\begin{tabular}[c]{@{}l@{}}Challenge \\ input \\sampled \\ equals to point \\ (5/10 tests)\end{tabular}} & \multicolumn{1}{l|}{2.92\%} & \multicolumn{1}{l|}{-} \\ \cline{3-5} 
\multicolumn{1}{|c|}{} & \multicolumn{1}{l|}{} & \multicolumn{1}{l|}{\begin{tabular}[c]{@{}l@{}}Challenge \\ input \\ sampled \\ does not  equal \\to point\\  (5/10 tests)\end{tabular}} & \multicolumn{1}{l|}{-} & \multicolumn{1}{l|}{0.789\%} \\ \hline
\end{tabular}
\end{center}
\vspace{-2mm}
\end{table}

\begin{table}[hbtp]
\caption{The average true and false positive rates for the correctness tests from the ibmq\_montreal quantum computer, where only phase flip errors are detected. Each test size is 10.}
\begin{center}
\begin{tabular}{lllll}
\hline
\multicolumn{3}{|c|}{\textbf{Test}} & \multicolumn{2}{c|}{\textbf{\begin{tabular}[c]{@{}c@{}}Average Program Results \\for Phase Flip Detection\\ (ibmq\_montreal)\end{tabular}}} \\ \hline
\multicolumn{1}{|c|}{\textit{\textbf{\#}}} & \multicolumn{2}{c|}{\textit{\textbf{Description}}} & \multicolumn{1}{c|}{\textit{\textbf{\begin{tabular}[c]{@{}c@{}}True Positive \\ (Correct \\ password \\ is accepted)\end{tabular}}}} & \multicolumn{1}{c|}{\textit{\textbf{\begin{tabular}[c]{@{}c@{}}False Positive\\ (Wrong \\ password \\ is accepted)\end{tabular}}}} \\ \hline
\multicolumn{1}{|l|}{1} & \multicolumn{2}{l|}{\begin{tabular}[c]{@{}l@{}}Challenge input \\ is equal to point\end{tabular}} & \multicolumn{1}{l|}{1.57\%}  & \multicolumn{1}{l|}{-} \\ \hline
\multicolumn{1}{|l|}{2} & \multicolumn{2}{l|}{\begin{tabular}[c]{@{}l@{}}Challenge input \\ has a permutation error\end{tabular}} & \multicolumn{1}{l|}{-} & \multicolumn{1}{l|}{1.66\%} \\ \hline
\multicolumn{1}{|l|}{3} & \multicolumn{2}{l|}{\begin{tabular}[c]{@{}l@{}}Challenge input \\ has an X error\end{tabular}} & \multicolumn{1}{l|}{-} & \multicolumn{1}{l|}{1.51\%} \\ \hline
\multicolumn{1}{|l|}{4} & \multicolumn{2}{l|}{\begin{tabular}[c]{@{}l@{}}Challenge input \\ has a Z error\end{tabular}} & \multicolumn{1}{l|}{-} & \multicolumn{1}{l|}{1.59\%} \\ \hline
\multicolumn{1}{|l|}{5} & \multicolumn{2}{l|}{\begin{tabular}[c]{@{}l@{}}Challenge input \\ has an X and Z error\end{tabular}} & \multicolumn{1}{l|}{-} & \multicolumn{1}{l|}{1.51\%} \\ \hline
\multicolumn{1}{|c|}{\multirow{2}{*}{6}} & \multicolumn{1}{l|}{\multirow{2}{*}{\begin{tabular}[c]{@{}l@{}}Challenge \\ input \\ sampled\\ from \\ challenge \\ distribution\end{tabular}}} & \multicolumn{1}{l|}{\begin{tabular}[c]{@{}l@{}}Challenge \\ input \\ sampled \\ equals to point \\ (5/10 tests)\end{tabular}} & \multicolumn{1}{l|}{1.57\%} & \multicolumn{1}{l|}{-} \\ \cline{3-5} 
\multicolumn{1}{|c|}{} & \multicolumn{1}{l|}{} & \multicolumn{1}{l|}{\begin{tabular}[c]{@{}l@{}}Challenge \\ input \\ sampled \\ does not equal \\ to point\\  (5/10 tests)\end{tabular}} & \multicolumn{1}{l|}{-} & \multicolumn{1}{l|}{1.62\%} \\ \hline
\end{tabular}
\end{center}
\vspace{-2mm}
\end{table}

\begin{table}[hbtp]
\caption{The average true and false positive rates for the correctness tests from the ibmq\_montreal quantum computer, where all errors are detected. Each test size is 10.}
\begin{center}
\begin{tabular}{lllll}
\hline
\multicolumn{3}{|c|}{\textbf{Test}} & \multicolumn{2}{c|}{\textbf{\begin{tabular}[c]{@{}c@{}}Average Program Results for \\ Bit and Phase Flip Detection \\ (ibmq\_montreal)\end{tabular}}} \\ \hline
\multicolumn{1}{|c|}{\textit{\textbf{\#}}} & \multicolumn{2}{c|}{\textit{\textbf{Description}}} & \multicolumn{1}{c|}{\textit{\textbf{\begin{tabular}[c]{@{}c@{}}True Positive \\ (Correct \\ password \\ is accepted)\end{tabular}}}} & \multicolumn{1}{c|}{\textit{\textbf{\begin{tabular}[c]{@{}c@{}}False Positive\\ (Wrong \\ password \\ is accepted)\end{tabular}}}} \\ \hline
\multicolumn{1}{|l|}{1} & \multicolumn{2}{l|}{\begin{tabular}[c]{@{}l@{}}Challenge input \\ is equal to point\end{tabular}} & \multicolumn{1}{l|}{0.0366\%}  & \multicolumn{1}{l|}{-} \\ \hline
\multicolumn{1}{|l|}{2} & \multicolumn{2}{l|}{\begin{tabular}[c]{@{}l@{}}Challenge input \\ has a permutation error\end{tabular}} & \multicolumn{1}{l|}{-} & \multicolumn{1}{l|}{0.0269\%} \\ \hline
\multicolumn{1}{|l|}{3} & \multicolumn{2}{l|}{\begin{tabular}[c]{@{}l@{}}Challenge input \\ has an X error\end{tabular}} & \multicolumn{1}{l|}{-} & \multicolumn{1}{l|}{0.0232\%} \\ \hline
\multicolumn{1}{|l|}{4} & \multicolumn{2}{l|}{\begin{tabular}[c]{@{}l@{}}Challenge input \\ has a Z error\end{tabular}} & \multicolumn{1}{l|}{-} & \multicolumn{1}{l|}{0.0415\%} \\ \hline
\multicolumn{1}{|l|}{5} & \multicolumn{2}{l|}{\begin{tabular}[c]{@{}l@{}}Challenge input \\ has an X and Z error\end{tabular}} & \multicolumn{1}{l|}{-} & \multicolumn{1}{l|}{0.0269\%} \\ \hline
\multicolumn{1}{|c|}{\multirow{2}{*}{6}} & \multicolumn{1}{l|}{\multirow{2}{*}{\begin{tabular}[c]{@{}l@{}}Challenge \\ input \\ sampled\\ from \\ challenge \\ distribution\end{tabular}}} & \multicolumn{1}{l|}{\begin{tabular}[c]{@{}l@{}}Challenge \\ input \\ sampled \\ equals to point \\ (5/10 tests)\end{tabular}} & \multicolumn{1}{l|}{0.0391\%} & \multicolumn{1}{l|}{-} \\ \cline{3-5} 
\multicolumn{1}{|c|}{} & \multicolumn{1}{l|}{} & \multicolumn{1}{l|}{\begin{tabular}[c]{@{}l@{}}Challenge \\ input \\ sampled \\ does not equal \\ to point\\  (5/10 tests)\end{tabular}} & \multicolumn{1}{l|}{-} & \multicolumn{1}{l|}{0.00977\%} \\ \hline
\end{tabular}
\end{center}
\vspace{-2mm}
\end{table}

\begin{figure*}
% \resizebox{\columnwidth}{!}
% {
\includegraphics[width=\textwidth,height=\textheight,keepaspectratio]{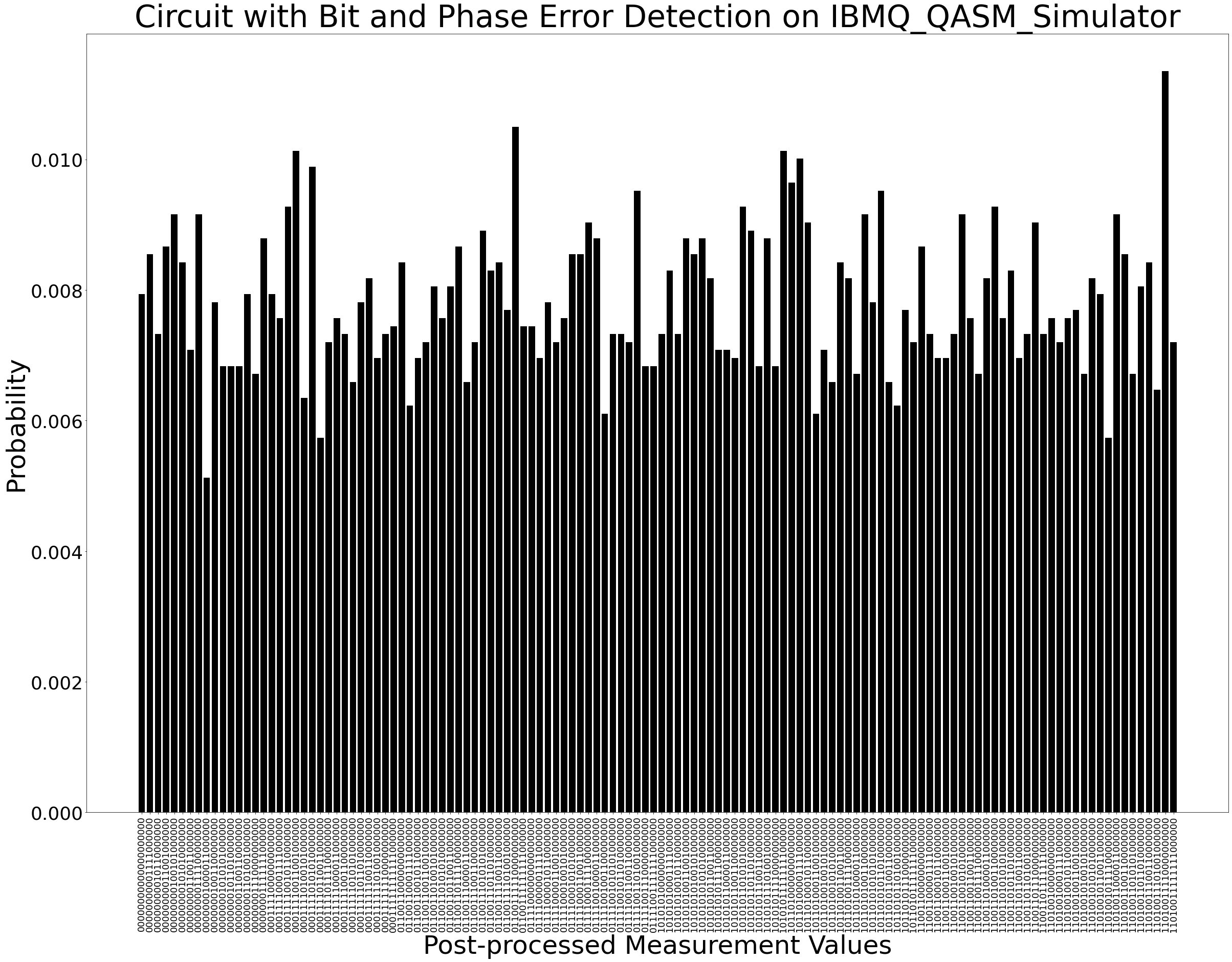}
\caption{The error syndrome measurement outcomes of a copy-protected  program for test one. The program is run on the ibmq\_qasm\_simulator. The measurement values have all been post-processed, and the sum of all the probabilities in the graph is equal to 1. The circuit was executed and sampled 8192 times.}
\end{figure*}

The measurement outcomes with bit or phase flip errors are not shown in Fig. 7 or Fig. 8 for better visibility of the results. Since the measurement values in the figures are post-processed, the error-free syndrome for detecting phase flips is all 0s. Additionally, the other 14 qubits are representative of the codewords of the encoded program for detecting bit flip errors.

\begin{figure}[htbp]
\resizebox{\columnwidth}{!}
{
\includegraphics{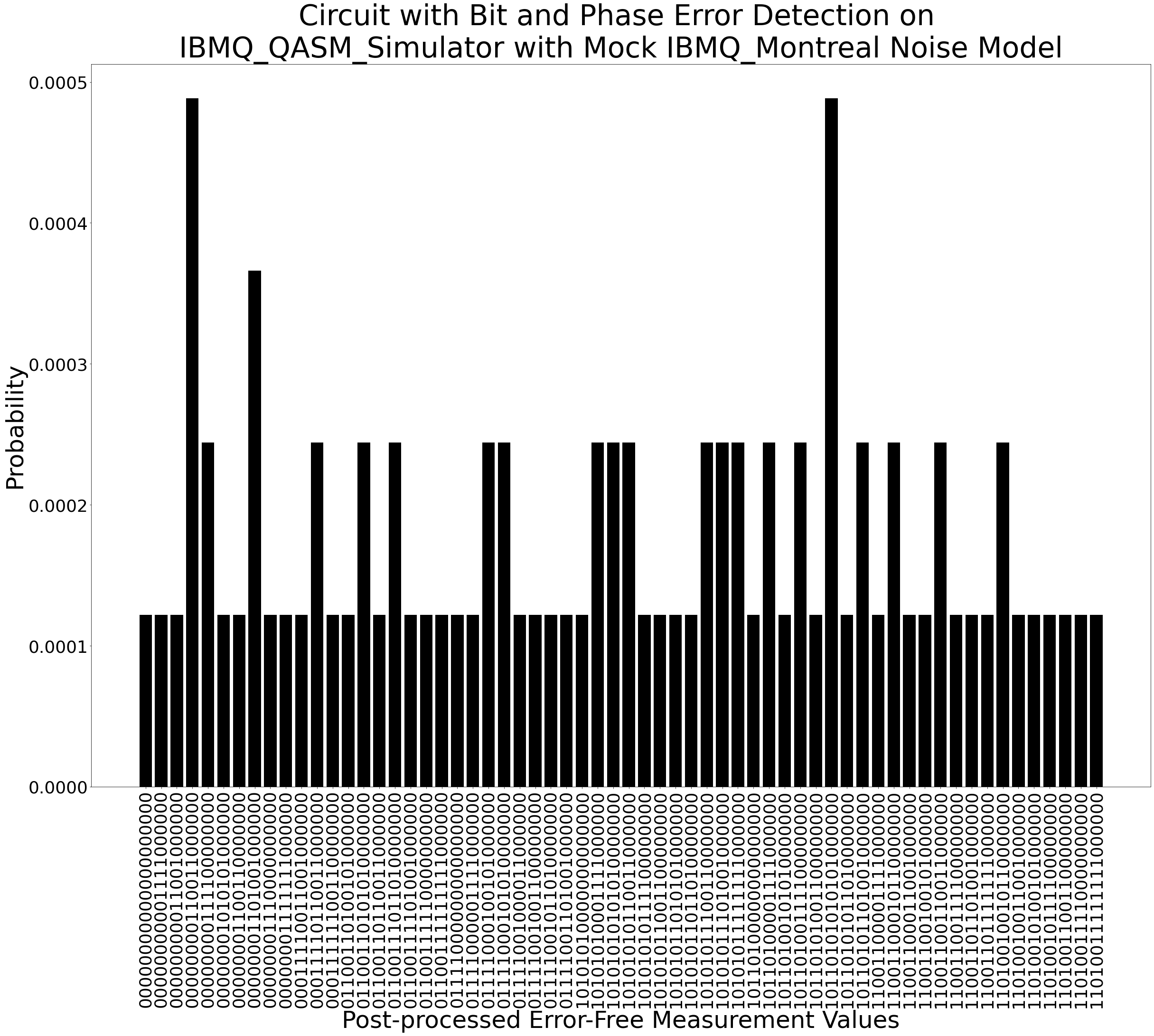}
}
\caption{The error syndrome measurement of a copy-protected program for test one. The program is run on the ibmq\_qasm\_simulator with a mock ibmq\_montreal device noise model. The measurement values have all been post-processed, and only the error-free measurement values are shown. The circuit was executed 8192 times, and the probability for obtaining the error-free measurement values are low.}
\label{fig}
\end{figure}

\begin{figure}[htbp]
\resizebox{\columnwidth}{!}
{
\includegraphics{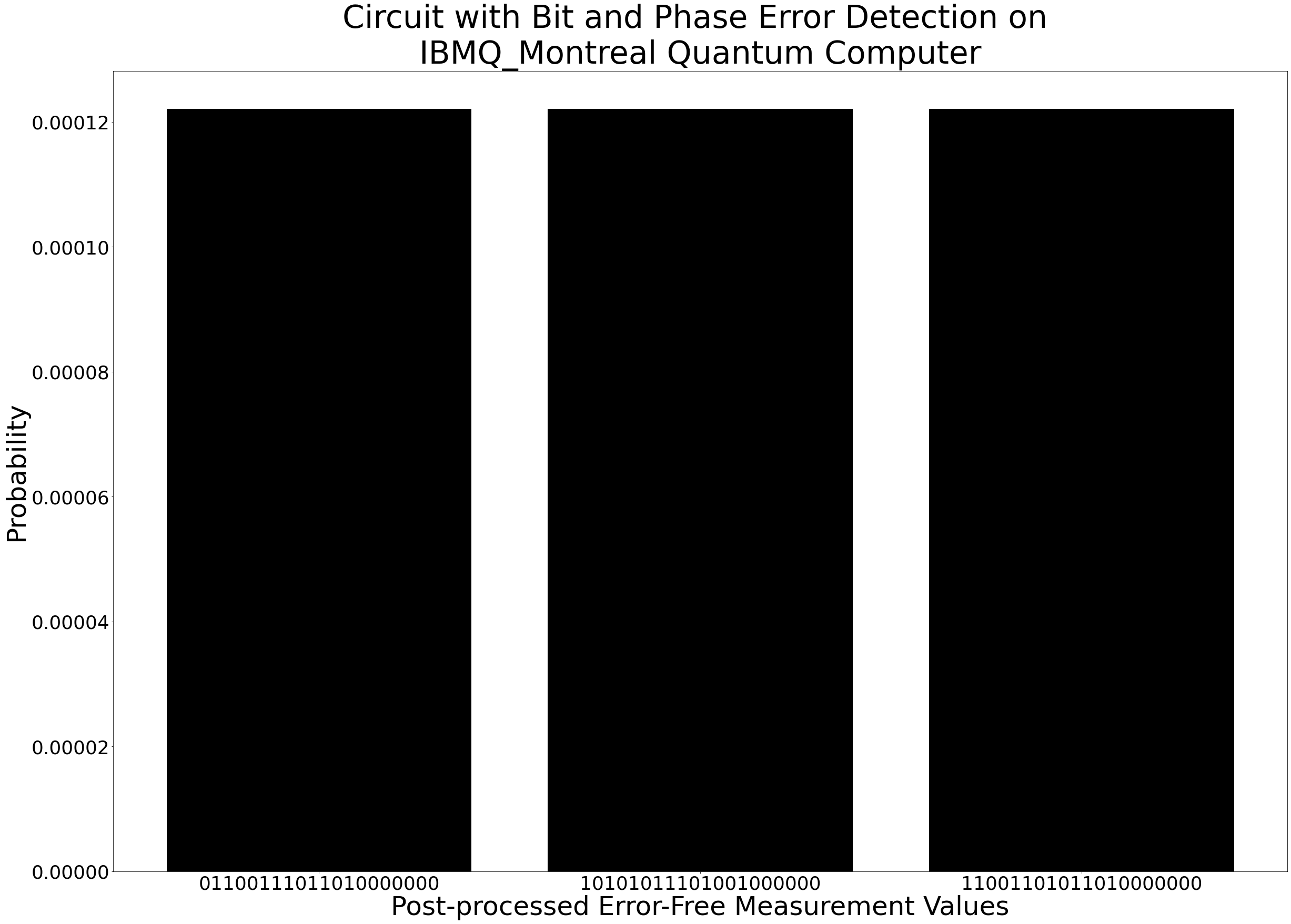}
}
\caption{The error syndrome measurement of a copy-protected  program for test one. The program is run on the ibmq\_montreal quantum device. The measurement values have all been post-processed, and only the error-free measurement values are shown. The circuit was executed 8192 times, and the probability for obtaining the error-free measurement values are low.}
\label{fig}
\end{figure}

\section*{Acknowledgments}

We would like to acknowledge CMC Microsystems for facilitating this research, specifically through their member access to the IBM Quantum Hub at Institut quantique. We are grateful to  Udson Mendes from CMC Microsystems, for invaluable work, including discussions on noise mitigation.

This work is supported in part by the Air Force Office of Scientific Research under award number FA9550-20-1-0375, Canada's NFRF, Canada's NSERC, an Ontario ERA, and the University of Ottawa's Research Chairs program.


\begin{thebibliography}{00}
\bibitem{b1} L. K. Grover, ``A fast quantum mechanical algorithm for database search,'' in \textit{annual ACM symposium Theory Comput.}, Jul. 1996, pp. 212-219, doi: \href{https://doi.org/10.1145/237814.237866}{10.1145/237814.237866}.
\bibitem{b2} 
A. Broadbent, S. Jeffery, S. Lord, S. Podder, and A. Sundaram, “Secure Software Leasing 
Without Assumptions,” in \textit{19th Theory of Cryptography Conference}, Nov. 2021, vol. 13042, pp. 90-120, doi:\href{https://link.springer.com/chapter/10.1007%2F978-3-030-90459-3_4}{	10.1007/978-3-030-90459-3\_4}.
\bibitem{b3} E. Alkim \textit{et al}, “FrodoKEM Learning With Errors Key Encapsulation Algorithm Specifications And Supporting Documentation,” Sept. 30, 2020. [Online]. Available: \href{https://frodokem.org/files/FrodoKEM-specification-20200930.pdf}{https://frodokem.org/files/FrodoKEM-specification-20200930.pdf}
\bibitem{b4}  S. K. Rao, D. Mahto, D. K. Yadav, and D. A. Khan, ``The AES-256 Cryptosystem Resists Quantum Attacks,'' \textit{Int. J. Adv. Res. Comput. Sci.}, vol. 8, pp. 404-408, Apr. 2017.
\bibitem{b5} HL. Huang \textit{et al.}, ``Homomorphic encryption experiments on IBM’s cloud quantum computing platform,'' \textit{Front. Phys.}, vol. 12:120305, Dec. 2017, doi: \href{https://doi.org/10.1007/s11467-016-0643-9}{10.1007/s11467-016-0643-9}. 
\bibitem{b6} A. Dash, S. Rout, B. K. Behera, and P. K. Panigrahi, ``Quantum Locker Using a Novel Verification Algorithm and Its Experimental Realization in IBM Quantum Computer,'' 2017, \textit{\href{https://arxiv.org/abs/1710.05196}{arXiv:1710.05196v2}}.
\bibitem{b7} D. Joy, M. Sabir, B. K. Behera, and P. K. Panigrahi, ``Implementation of quantum secret sharing and quantum binary voting protocol in the IBM quantum computer,'' \textit{Quantum Inf. Process.}, vol. 19, no. 33, Dec. 2019, doi: \href{https://doi.org/10.1007/s11128-019-2531-z}{10.1007/s11128-019-2531-z}.
\bibitem{b8}Y. Dulek and F. Speelman, ``Quantum ciphertext authentication and key recycling with the trap code,'' in \textit{13th Conference Theory Quantum Comput., Commun. and Crypto.---TQC 2018}, 2018, pp. 1-17, doi: \href{https://drops.dagstuhl.de/opus/volltexte/2018/9248/}{10.4230/LIPIcs.TQC.2018.1}. 
\bibitem{b9} S. D. Buchbinder, C. L. Huang, and Y. S. Weinstein, ``Encoding an arbitrary state in a [7,1,3] quantum error correction code,'' \textit{Quantum Inf. Process.}, vol. 12, pp. 699–719, Feb. 2013, doi: \href{https://doi.org/10.1007/s11128-012-0414-7}{10.1007/s11128-012-0414-7}.
\bibitem{b10} A.Broadbent, “Delegating Private Quantum Computations,” \textit{Canadian J. Phys.}, vol. 93, no. 9, pp. 941-946, Jun. 2015, doi: \href{https://doi.org/10.1139/cjp-2015-0030}{10.1139/cjp-2015-0030}.
\bibitem{b11} IBM Quantum Computing, “System Properties,” 2021. [Online]. Available: \href{https://quantum-computing.ibm.com/lab/docs/iql/manage/systems/properties}{https://quantum-computing.ibm.com/lab/docs/iql/manage/systems/\\properties}.
\end{thebibliography}
\end{document}